\documentclass[12pt,preprint]{aastex}               
\shortauthors{D.~A.~Uzdensky}

\begin{document}

\title{On the Axisymmetric Force-Free Pulsar Magnetosphere
\thanks{preprint NSF-KITP-03-36}}
\author{Dmitri A. Uzdensky}
\affil{Kavli Institute for Theoretical Physics, University of California} 
\affil{Santa Barbara, CA 93106}
\email{uzdensky@kitp.ucsb.edu}
\date{May 15, 2003}

\begin{abstract}

We investigate the axisymmetric magnetosphere of an aligned 
rotating magnetic dipole surrounded by an ideal force-free 
plasma. We concentrate on the magnetic field structure around 
the point of intersection of the separatrix between the open 
and closed field-line regions and the equatorial plane. We 
first study the case where this intersection point is located 
at the Light Cylinder. We find that in this case the separatrix 
equilibrium condition implies that all the poloidal current must
return to the pulsar in the open-field region, i.e., that there 
should be no finite current carried by the separatrix/equator 
current sheet. We then perform an asymptotic analysis of the pulsar 
equation near the intersection point and find a unique self-similar 
solution; however, a Light Surface inevitably emerges right outside 
the Light Cylinder. We then perform a similar analysis for the situation 
where the intersection point lies somewhere inside the Light Cylinder, 
in which case a finite current flowing along the separatrix and the 
equator is allowed. We find a very simple behavior in this case, 
characterized by a 90-degree angle between the separatrix and 
the equator and by finite vertical field in the closed-field 
region. Finally, we discuss the implications of our results for 
global numerical studies of pulsar magnetospheres.
\end{abstract}



\section{Introduction}
\label{sec-intro}

The structure of the pulsar magnetosphere has been an active
area of research for more that 30 years, starting with the 
pioneering work by Goldreich \& Julian (1969). Despite the
fact that real astrophysical radio-pulsars are believed to 
be oblique rotating magnetic dipoles, much of the theoretical
effort has been devoted to a significantly simpler case of 
an aligned rotating magnetic dipole (i.e., a ``non-pulsing
pulsar''), in which case the problem becomes stationary and
axisymmetric. The situation becomes even more tractable 
if one also assumes the space around the pulsar to be 
filled with a plasma that, on one hand, is dense enough
to shorten out the longitudinal electric fields, thus 
providing the basis for the ideal-magnetohydrodynamic
approximation, and, on the other hand, is at the same 
time tenuous enough for all the non-electromagnetic 
forces to be negligible, thus enabling one to regard 
the plasma as force-free. 

All these simplifications have lead early on to the derivation, 
simultaneously by several researchers, of the main equation governing 
the structure of the magnetosphere, the so-called pulsar equation 
(Scharlemann \& Wagoner~1973; Michel~1973b; Okamoto~1974), which is 
essentially a special-relativistic generalization of the well-known 
force-free Grad--Shafranov equation. This equation is a quasi-linear 
elliptic second-order partial differential equation with a regular 
singular surface, the so-called Light Cylinder. Despite the fact that 
this, in general non-linear, equation was first derived thirty years 
ago, most of the attempts to solve it, both analytical and numerical, 
have been limited, until very recently, to its linear special cases 
and to the region inside the Light Cylinder (e.g., Scharlemann \& 
Wagoner~1973; Michel~1973a; Beskin et al.~1983; Beskin \& Malyshkin~1998).
These self-imposed restrictions can of course be attributed to the 
considerations of mathematical convenience and tractability; however, 
they are not very well physically motivated and thus may be too 
restrictive to be relevant to real astrophysical systems.

At the same time, by the late 1990s, the available computing
resources and numerical techniques have become sufficiently
powerful for the most general nonlinear problem, involving 
the regions on both sides of the Light Cylinder, to become
tractable. In their pioneering work, Contopoulos~et~al. (1999, 
hereafter CKF) have obtained the first (and apparently unique)
numerical solution of the general problem, albeit with a
rather poor spatial resolution. Their approach has later
been used by Ogura \& Kojima (2003) who have obtained 
essentially the same results, but with a higher numerical 
resolution. Both of these groups have focussed on the global 
structure of the solution; consequently, they have not paid 
enough attention, in our opinion, to some key but subtle 
issues regarding the separatrix between the open and closed 
field-line regions and, especially, the point of intersection 
of this separatrix with the equatorial plane. As a result,
the separatrix has not, we believe, been treated correctly,
and, in particular, the separatrix equilibrium condition
(Okamoto~1974; Lyubarskii~1990) has not been satisfied 
close to the intersection point. We thus suspect that 
their solution is probably not quite correct.

In this paper we try to clarify how the magnetic field near 
the separatrix and especially around the separatrix--equator
intersection point should be treated, as this question appears
to be extremely important for setting up the {\it correct}
boundary conditions for the global problem.

In \S~\ref{sec-basic} we review the basic equations describing the
ideal-MHD force-free magnetosphere of an aligned rotating dipole;
we pay special attention to the so-called Light Cylinder regularity
condition and discuss how this condition could be used to find the
correct form of the poloidal-current function and, simultaneously,
to obtain the solution that passes smoothly across the Light Cylinder.
Then, in \S~\ref{sec-Y-point}, we investigate some general properties
of the magnetic field near the putative separatrix between the open
and closed field regions; we use the separatrix equilibrium condition
to show that, if the separatrix intersects the equator at the Light 
Cylinder (as has been assumed in the numerical simulations by CKF
and by Ogura \& Kojima~2003), then the poloidal current and hence 
the toroidal magnetic field have to vanish on the last open field 
line above the separatrix. We then use these findings in \S~\ref
{sec-analysis} to construct a unique self-similar asymptotic solution 
of the pulsar equation in the vicinity of such an intersection point;
in particular, we find all the power-law exponents describing the field
near this point and also the angle between the separatrix and the equator.
At the end of this section we show that a Light Surface (a surface where
the electric field becomes equal to the magnetic field and therefore
where the particle drift velocity reaches the speed of light) has to 
appear right outside the Light Cylinder (it starts at the intersection
point and extends outward at a finite angle with respect to the equator).
We thus conclude that the only way to get a force-free solution that 
would be valid at least some finite distance beyond the Light Cylinder, 
is to consider the case of a separatrix--equator intersection point 
lying some finite distance inside the Light Cylinder. We investigate 
the magnetic field structure around such a point in \S~\ref{sec-T-point}.
We find that in this case the poloidal-current function may stay finite,
but its derivative with respect to the poloidal flux has to go to zero 
on the last open field line; we then find the asymptotic behavior of 
the magnetic field around the intersection point corresponding to 
this case. Finally, we present our conclusions in \S~\ref{sec-conclusions},
where we also discuss the implications of our results for the past and
future numerical studies of the axisymmetric pulsar magnetospheres.


\section{Basic Equations}
\label{sec-basic}

We consider the magnetosphere of an aligned rotating magnetic
dipole under the assumptions of stationarity, axial symmetry
with respect to the rotation axis~$Z$, and reflection symmetry 
with respect to the equator $Z=0$. We shall include the effects 
of special relativity but will ignore general-relativistic effects 
(i.e., work in Euclidean space). We shall assume that the dipole 
is surrounded by a very tenuous, but highly conducting plasma.
The first of these assumptions enables us to neglect all non-electromagnetic
forces (i.e., gravity, inertial, and pressure forces), and thus
to conclude that the structure of the magnetosphere is governed 
by the relativistic force-free equation:
\begin{equation}
\rho {\bf E} + {{\bf j\times B}\over c} = 0 \, .
\label{eq-force-free}
\end{equation}
Here, the electric charge density $\rho$ and the electric current 
density ${\bf j}$ are related to the electric and magnetic field
through the steady-state Maxwell equations:
\begin{eqnarray}
\rho &=& {1\over{4\pi}}\, \nabla \cdot {\bf E}\, , 
\label{eq-Maxwell-rho}                 \\
{\bf j} &=& {c\over{4\pi}}\,  \nabla\times{\bf B} \, .
\label{eq-Maxwell-j}
\end{eqnarray}

The second of our assumptions concerning the magnetospheric plasma,
i.e., the assumption of infinite conductivity, enables us to use
ideal magnetohydrodynamics (MHD):
\begin{equation}
{\bf E} + {{{\bf v}\times{\bf B}}\over c} = 0 \, .
\label{eq-ideal-MHD}
\end{equation}

These equations (with the appropriate boundary conditions) should
be sufficient for determining the structure of the magnetosphere.

As is well known, an axisymmetric magnetic field ${\bf B}$ can
be described in terms of two functions, the poloidal magnetic 
flux (per one radian in the azimuthal direction) $\Psi$ and
the poloidal current~$\tilde{I}$, as
\begin{equation}
{\bf B} = \nabla\Psi\times\nabla\phi + \tilde{I} \nabla\phi \, ,
\label{eq-B}
\end{equation}
where $\phi$ is the azimuthal (or toroidal) angle.
In cylindrical coordinates $(R,\phi,Z)$ the magnetic 
field components then are:
\begin{eqnarray}
B_R &=& -{1\over R}\, {\partial\Psi\over{\partial Z}}\, , 
\label{eq-B_R-1}                                         \\
B_Z &=& {1\over R}\, {\partial\Psi\over{\partial R}}\, , 
\label{eq-B_Z-1}                                         \\
B_\phi &=& {\tilde{I}\over R}\, . 
\label{eq-B_phi-1}
\end{eqnarray}

Next, by applying Faraday's law $\partial{\bf B}/\partial t = 
-c\nabla\times{\bf E}$ in a steady state, we see that 
$\nabla\times{\bf E}=0$, and hence ${\bf E}=-\nabla\Phi$; 
together with the assumption of axial symmetry, this gives 
\begin{equation}
E_\phi = 0 \, ,
\label{eq-E_phi=0}
\end{equation}
that is, the electric field is purely poloidal. Since, as it follows
from equation~(\ref{eq-force-free}) or~(\ref{eq-ideal-MHD}), the electric 
field must be perpendicular to the magnetic field, we can write it as
\begin{equation}
{\bf E} = {R\Omega\over c}\ {\bf B}_{\rm pol}\times\hat{\phi} =
-\, {\Omega\over c}\, \nabla\Psi \, ,
\label{eq-E}
\end{equation}
where ${\bf B}_{\rm pol} = \nabla\Psi\times\nabla\phi$ is 
the poloidal magnetic field, and $\hat{\phi}$ is the unit 
vector in the $\phi$-direction.
By substituting this relationship into $\nabla\times{\bf E}=0$,
we find that~$\Omega$, which has the meaning of the angular 
velocity of magnetic field lines, is constant along the field 
lines, $\Omega=\Omega(\Psi)$. This is the well-known Ferraro 
isorotation law.

In the problem we are interested in, all the field lines are tied,
at least at one end, to the surface of the pulsar, which is assumed
to be rotating as a solid body with some uniform angular velocity~$\Omega_*$.
We also assume that our ideal-MHD assumption is valid all the way up to
the pulsar surface, i.e., that there is no appreciable gap between this 
surface and the ideal-MHD region. Therefore, from now on, we shall assume 
that all the field lines rotate with the same angular velocity%
\footnote{The case where $\Omega(\Psi)\neq\Omega_*$ was considered, e.g.,
by Beskin~et~al. (1983) and by Beskin \& Malyshkin (1998).}
\begin{equation}
\Omega(\Psi) = \Omega_* = {\rm const} \, .
\label{eq-Omega=Omega_*}
\end{equation}

From equation~(\ref{eq-E}) we find
\begin{equation}
|{\bf E}| = |{\bf B}_{\rm pol}|\, x\, ,
\label{eq-|E|}
\end{equation}
where
\begin{equation}
x \equiv {R\Omega\over c} = {R\over{R_{LC}}} 
\label{eq-def-x}
\end{equation}
is the cylindrical radius normalized to the radius of the 
Light Cylinder (LC) $R_{LC}\equiv c/\Omega$. Thus, we see 
that $|{\bf E}| < |{\bf B}_{\rm pol}|$ inside the LC and 
$|{\bf E}| > |{\bf B}_{\rm pol}|$ outside the~LC.

Now let us turn to the force-free equation~(\ref{eq-force-free}). 
The toroidal component of this equation, together with equation~(\ref
{eq-E_phi=0}), gives:
\begin{equation}
{\bf j}_{\rm pol} \times {\bf B}_{\rm pol} = 0 \quad \Rightarrow \quad
[\nabla\tilde{I}\times\nabla\phi] \times [\nabla\Psi\times\nabla\phi] = 0
\quad \Rightarrow \quad \tilde{I}=\tilde{I}(\Psi) \, ,
\label{eq-I=I(Psi)}
\end{equation}
and the poloidal component (perpendicular to~${\bf B}_{\rm pol}$)
of equation~(\ref{eq-force-free}) can be 
written as
\begin{equation}
(1-x^2)\, \biggl[ {{\partial^2\Psi}\over{\partial x^2}} +
{{\partial^2\Psi}\over{\partial z^2}} \biggr] -
{{1+x^2}\over x}\, {\partial\Psi\over{\partial x}} = - I I'(\Psi) \, ,
\label{eq-pulsar}
\end{equation}
where we have introduced $z\equiv Z/R_{LC}$ and where we also
incorporated $R_{LC}$ into $I(\Psi)$ on the right-hand side (RHS)
by defining
\begin{equation}
I(\Psi) \equiv R_{LC} \tilde{I}(\Psi) \, .
\label{eq-I-I_tilde}
\end{equation}

Then the magnetic field components can be rewritten in terms
of~$\Psi(x,z)$ as
\begin{eqnarray}
B_R &=& -\, {1\over{x R_{LC}^2}}\, {\partial\Psi\over{\partial z}}\, , 
\label{eq-B_R-2}                                         \\
B_Z &=& {1\over{x R_{LC}^2}}\, {\partial\Psi\over{\partial x}}\, , 
\label{eq-B_Z-2}                                         \\
B_\phi &=& {{I(\Psi)}\over{x R_{LC}^2}}\, .  
\label{eq-B_phi-2}
\end{eqnarray}

Equation (\ref{eq-pulsar}) is the famous {\it pulsar equation} 
(Scharlemann \& Wagoner~1973; Michel~1973b; Okamoto~1974), also 
known in literature as the relativistic force-free Grad--Shafranov 
equation (e.g., Beskin~1997). It is an elliptic second-order partial 
differential equation (PDE) for the flux function~$\Psi(x,z)$; the 
left-hand side (LHS) of this equation is linear, whereas the RHS is, 
in general, nonlinear.

One very important feature of equation~(\ref{eq-pulsar}) is that 
it has a regular singular surface at the Light Cylinder~$x=1$.
Both indicial (or characteristic) exponents for this equation 
are equal to zero and so a general solution of equation~(\ref
{eq-pulsar}) diverges logarithmically at this surface (e.g., 
Bender \& Orszag~1978). On physical grounds, one imposes an 
additional condition that both the function~$\Psi(x,z)$ and 
its 1st and 2nd derivatives remain finite near the~LC. For 
any given function $I(\Psi)$, it is indeed possible to find 
such non-divergent solutions separately on each side of the~LC. 
Then, as can be seen from equation~(\ref{eq-pulsar}) itself, 
the contribution of the first term on the LHS [the term 
proportional to $(1-x^2)$] vanishes as $x\rightarrow 1$, 
and thus such non-divergent solutions on both sides of 
the~LC have to satisfy the well-known LC regularity 
condition (e.g., Scharlemann \& Wagoner~1973; Okamoto~1974; 
Beskin~1997):
\begin{equation}
{{\partial\Psi}\over{\partial x}}\, (x=1,z) = {1\over 2}\, I I'(\Psi) \, .
\label{eq-regularity-LC}
\end{equation}

This condition is very important and merits a few extra words 
of discussion. Our force-free equation with a constant field-line
angular velocity~$\Omega$ is a 2nd-order equation with one, a priori 
unknown, integral of motion, $I(\Psi)$, and one singular surface, the~LC. 
According to general theory (Beskin~1997), one then needs two boundary 
conditions to set up the problem properly. We believe that the correct 
approach here is to set the boundary conditions for the function~$\Psi$ 
at both the inner boundary (inside the~LC: on the surface of the star, 
the symmetry axis, the separatrix between the open and closed field-line 
regions, and maybe the equator) and at whatever outer boundary (outside 
the~LC: the equator and infinity or the Light Surface). At the same time 
the poloidal current function~$I(\Psi)$ should not be prescribed explicitly 
at any boundary of the domain; instead, it is to be determined from the 
matching condition at the singular surface, i.e., the~LC (see below).

Thus, our position differs from that of Beskin (see Beskin et al.~1983,
and Beskin~1997) who proposed prescribing~$I(\Psi)$ explicitly at the 
surface of the pulsar. Indeed, if one considers the region inside the 
LC only, then, upon prescribing both~$\Psi$ and~$I(\Psi)$ on the inner 
boundary (the pulsar surface, etc.), one should be able to obtain a 
solution that is regular at the~LC by using the regularity condition~(\ref
{eq-regularity-LC}) at the other boundary of this inner region, i.e., 
at the~LC. Thus, the solution in this inner region is then completely 
determined; as such, it is totally independent of what happens outside 
the~LC. This does not seem physical; indeed, one should then be able to 
take this solution and continue it smoothly across the~LC, thus prescribing 
both $\Psi(1,z)$ and $\Psi_x(1,z)$ at the outer side of the~LC. But then 
the problem of finding a solution in the region outside the LC becomes 
over-determined, as we now have two conditions at $x=1$, in addition to 
any conditions at the outer boundary. For example, in the very important 
particular case of the domain under consideration extending all the way 
to radial infinity, one cannot actually prescribe any specific boundary 
conditions at infinity because equation~(\ref{eq-pulsar}) has a regular 
singularity there. Instead, however, one imposes a regularity condition 
at infinity. In spherical polar coordinates%
\footnote
{Note that in the rest of the paper we shall use a totally different 
set of polar coordinates defined in the vicinity of the separatrix
intersection point, for which we shall use the same notation ($r,\theta$). 
We hope that this does not cause any confusion.}
($r,\theta,\phi$) this regularity condition can be written as
\begin{equation}
\Psi_{\theta\theta}+\Psi_{\theta} \cot\theta = 
{{II'(\Psi)}\over{\sin^2\theta}}, \qquad r\rightarrow\infty \, .
\label{eq-regularity-infinity}
\end{equation}
This regularity condition has a very simple physical meaning: 
it is the condition of the force balance in the $\theta$-direction
between the toroidal magnetic field~$B_{\phi}$ and the poloidal
electric field~$E_\theta$ (which both become much larger than 
the poloidal magnetic field $B_{\rm pol}\simeq B_r$ as one
approaches infinity). Thus, in this case the region outside
the LC does not have any boundary conditions at all (apart from the 
condition $\Psi={\rm const}$ at the equator), but has two regularity
conditions, one at the~LC, and the other at infinity. These two 
regularity conditions are already sufficient to uniquely determine
the solution $\Psi(x>1,z)$ for a given~$I(\Psi)$. Therefore, if, in 
addition to these two regularity conditions, one also tries to impose 
the function $\Psi(1,z)$ along the~LC, the system becomes over-determined. 
Thus we come to the conclusion that prescribing $I(\Psi)$ at the inner 
boundary is not appropriate. One could, however, try to use the outer 
boundary condition (or the regularity condition at infinity) as the 
condition that fixes~$I(\Psi)$. To do this, one first sets the inner 
boundary conditions, picks an initial guess for~$I(\Psi)$, and uses~(\ref
{eq-regularity-LC}) to obtain a regular solution inside the~LC; 
one then continues this solution smoothly across the~LC and solves 
equation~(\ref{eq-pulsar}) in the outer region as an initial-value 
problem (with both~$\Psi$ and $\partial_x\Psi$ specified at~$x=1$);
as a result, one gets a mismatch at the outer boundary between the 
obtained solution and the desired outer boundary condition [or, in 
the case of an infinite domain, one would presumably fail to get 
a convergent solution satisfying the regularity condition~(\ref
{eq-regularity-infinity})]. Then one iterates with respect to~$I(\Psi)$ 
until this outer-boundary mismatch is zero (or until a solution regular 
at infinity is achieved). In reality, however, such an approach may not 
be practical as it involves solving an initial-value problem for an 
elliptic equation, which is not a well-posed problem. 

Instead, we advocate the approach adopted by CKF. In this approach one 
considers the regions inside the~LC ($x<1$) and outside the~LC ($x>1$) 
separately. First, one makes an initial guess for~$I(\Psi)$ and prescribes 
the corresponding boundary conditions for~$\Psi$ at the inner boundary of 
the the region inside the~LC and at the outer boundary of the the region 
outside the~LC [or, if this outer boundary is at the radial infinity, 
imposes the regularity condition~(\ref{eq-regularity-infinity}) there]. 
Then, one uses, again separately in each region, the regularity 
condition~(\ref{eq-regularity-LC}) in lieu of a boundary condition 
at the~LC [in practice, one can think of~(\ref{eq-regularity-LC}) 
as of a mixed-type Dirichlet--von~Neumann boundary condition 
relating the value of~$\Psi$ at the surface~$x=1$ to the value 
of its derivative normal to this surface]. Thus one obtains two 
solutions, one inside and the other outside the~LC, that correspond 
to the same function~$I(\Psi)$ and are both regular at~$x=1$. However, 
these solutions also depend on their respective boundary or regularity 
conditions set at the inner/outer boundaries of the domain. Hence, in 
general, although they are both regular at the~LC, these solutions are 
not going to coincide at~$x=1$. The mismatch $\Delta\Psi_{LC}(z)$ depends 
on both the chosen function $I(\Psi)$ and on the inner and outer boundary 
conditions. One then iterates with respect to~$I(\Psi)$ in order to 
minimize this mismatch. We believe that for a given set of boundary 
conditions there should be only one choice of~$I(\Psi)$ for which the 
mismatch $\Delta\Psi_{LC}$ vanishes and hence the function~$\Psi$ becomes 
continuous along the entire~LC, i.e., $\Psi_{\rm in} (x\rightarrow 1,z)=
\Psi_{\rm out} (x\rightarrow 1,z)$. Once this special function~$I(\Psi)$ 
is found, it then follows from~(\ref{eq-regularity-LC}), which is 
satisfied separately on both sides of the~LC, that $\partial_x\Psi$ 
is also continuous, and so the entire solution passes smoothly across 
the~LC. Thus, one can say that the function $I(\Psi)$ is determined by 
the regularity condition ~(\ref{eq-regularity-LC}) applied separately 
on both sides of the~LC, plus the condition of matching of the two 
solutions. Another, equivalent way to put it, is to say that $I(\Psi)$ 
is determined by the (non-trivial!) requirement that the derivative 
$\partial_x\Psi(x=1,z)$ actually {\it exists} at the~LC [whereby 
$I(\Psi)$ is expressed in terms of this derivative via equation~(\ref
{eq-regularity-LC})].

This approach has been implemented successfully in the pioneering
numerical work by CKF and then subsequently by Ogura \& Kojima (2003).
Both these groups have used a relaxation procedure to arrive at $I(\Psi)$
that corresponded to a unique solution that was both continuous and 
smooth at the~LC.


\section{Behavior near the separatrix Y-point: general considerations}
\label{sec-Y-point}

\begin{figure}
\plotone{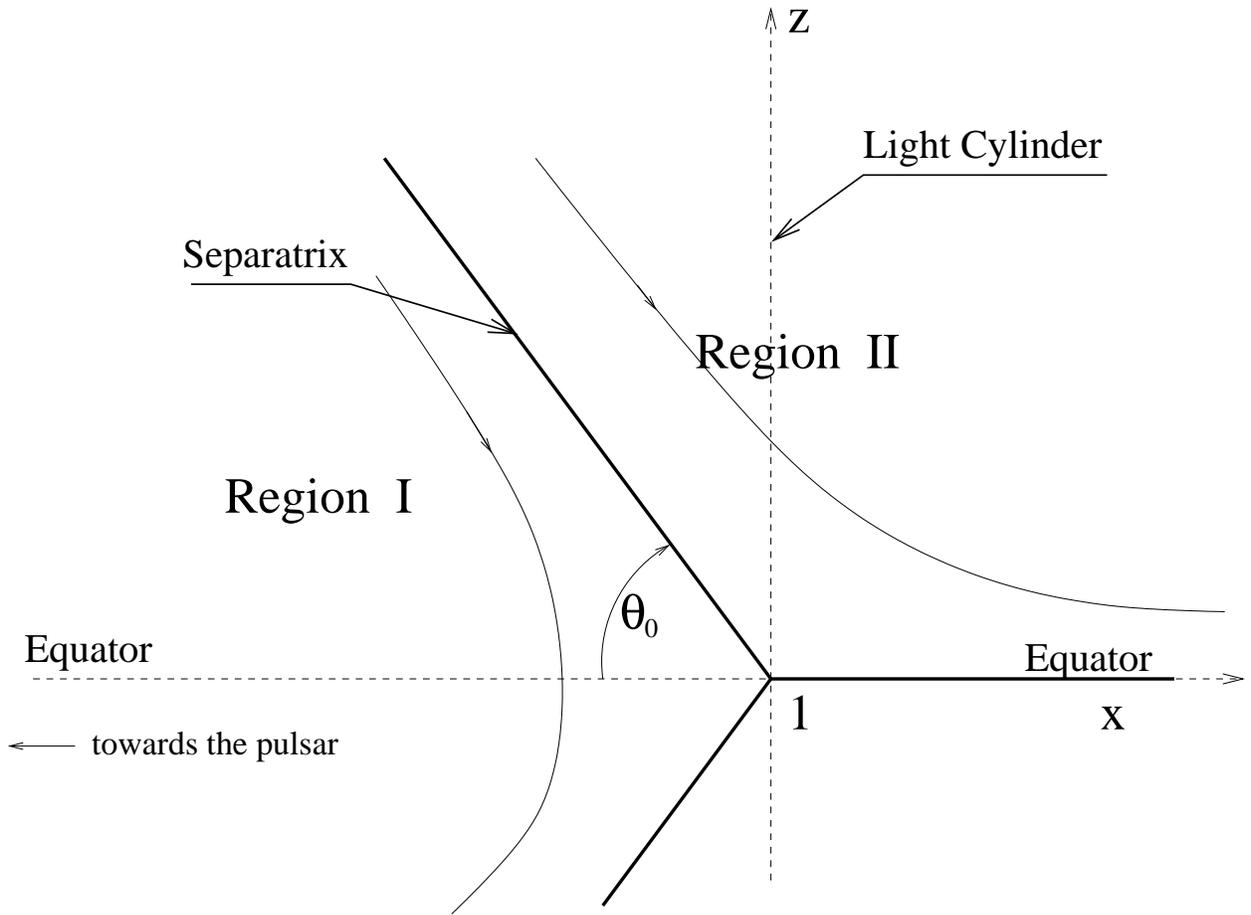}
\figcaption{A schematic drawing of the Y-type separatrix intersection
point in the case $x_0=1$.
\label{fig-geometry}}
\end{figure}

The main focus of this paper is the behavior of the magnetic field 
in the vicinity of the point of intersection of the separatrix $\Psi=
\Psi_s$ between the region of closed field lines (region~I) and the 
region of open field lines (region~II) with the equator $z=0$ (see
Fig.~\ref{fig-geometry}). We shall generally call this point the 
{\it separatrix intersection point}. Our interest in this non-trivial 
problem is fueled by the belief that understanding the key features of 
this behavior is absolutely crucial to devising the proper boundary 
conditions and providing verification benchmarks for any future global 
numerical investigations of a force-free pulsar magnetosphere. At the 
same time, the treatment of the magnetic field around this very special 
point is expected to require a certain degree of subtlety and delicacy, 
as noted, for example, by Beskin et~al. (1983) and by Ogura \& Kojima~(2003).

Among the questions that need to be answered are: what is the radial 
position $x_0$ of the intersection point (namely, whether this point 
lies at the~LC, $x_0=1$, of inside the~LC, $x_0<1$)? and what is the 
angle $\theta_0$ at which the separatrix approaches the equator at 
this point, the three a priori possibilities being $\theta_0=0$ (in 
which case we'll call this point the cusp-point), $0<\theta_0<\pi/2$ 
(the Y-point), and $\theta_0=\pi/2$ (the T-point)?

In our analysis, we shall assume that equations~(\ref{eq-force-free}) 
and~(\ref{eq-ideal-MHD}) that describe our system are valid almost 
everywhere in the vicinity of the separatrix intersection point, i.e., 
everywhere with perhaps the exception of measure-zero regions. Thus, 
we shall allow for the presence in our system of current sheets of 
infinitesimal thickness, across which the magnetic field can experience 
a finite jump. We limit our consideration to the situation where such 
current sheets can be present only along the separatrix between regions~I 
and~II for $x<x_0$ and along the equatorial ($z=0$) separatrix between
the upper and lower open-field regions for $x>x_0$. Thus, we have one 
current sheet that lies on the equator at $x>x_0$ and at $x=x_0$ splits
into two symmetrical current sheets (one in each hemisphere) lying along
the separatrix $\Psi=\Psi_s$ (see, e.g., Okamoto~1974). At the same time 
we assume that our equations apply everywhere else at least in some 
region around the intersection point, including the portion of this 
region that lies outside the LC (in the case $x_0=1$).

Let us first assume that the separatrix between closed and open
field-line regions reaches the~LC (i.e., $x_0=1$), and let us 
consider the separatrix equilibrium condition (e.g., Okamoto 1974;
Lyubarskii~1990). Indeed, whether or not the separatrix contains a 
current sheet (which we assume to be infinitesimally thin), it must 
satisfy the condition of force balance, which is obtained by integrating 
equation~(\ref{eq-force-free}) across the separatrix:%
\footnote
{Strictly speaking, equation~(\ref{eq-force-free}) may not be valid 
inside a current sheet as other forces may also be important in the 
pressure balance there. However, gravity and the plasma inertia are 
small because the separatrix current layer is assumed to be very thin; 
and the plasma pressure gradient, while not necessarily small inside 
the current layer, gives, when integrated across the layer, just the 
difference between the values of the pressure on both sides outside 
the separatrix, both of which are small by assumption.}
\begin{equation}
({\bf B}^2-{\bf E}^2)^I =({\bf B}^2-{\bf E}^2)^{II} \, .
\label{eq-sepx-equil-1}
\end{equation}

By using equation~(\ref{eq-|E|}) and also the fact that $B_\phi=0$ 
in region~I, we can rewrite this condition as 
\begin{equation}
[({\bf B}_{\rm pol}^I(l))^2-({\bf B}_{\rm pol}^{II}(l))^2]\, (1-x^2) =
(B_\phi^{II})^2 = {1\over{R_{LC}^4}}\, {I^2(\Psi_s)\over{x^2}} \, ,
\label{eq-sepx-equil-2}
\end{equation}
where $l$ marks the distance from the intersection point ($z=0$) 
along the separatrix.

Now, as the separatrix approaches the LC ($x\rightarrow 1$, 
$l\rightarrow 0$), we get
\begin{equation}
2(1-x)\, [({\bf B}_{\rm pol}^I(l))^2-({\bf B}_{\rm pol}^{II}(l))^2] = 
{{I_s^2}\over{R_{LC}^4}} \, ,
\label{eq-sepx-equil-endpt}
\end{equation}
where we have defined
\begin{equation}
I_s \equiv \lim\limits_{\Psi\rightarrow\Psi_s} \ I(\Psi<\Psi_s)
\label{eq-def-Is}
\end{equation}
as the value of the poloidal current of the last open field line~$\Psi_s$
(above the separatrix current sheet). Since $I_s$ is constant along this 
last open field line, and hence is independent of~$l$, we immediately see 
that in order to satisfy equation~(\ref{eq-sepx-equil-endpt}) in the limit 
$x\rightarrow 1$, we must have either:\\
{\it (i)} $I_s=0$, i.e., all poloidal current that flows out of 
the pulsar must return back to the pulsar along open magnetic field 
lines, with no finite poloidal current flowing along the separatrix;%
\footnote
{This scenario is in agreement with the findings by Okamoto~1974
and also with the particular case $\beta_0=0$ of the compatibility 
condition by Beskin~et~al. (1983) and Beskin \& Malyshkin (1998).} 
simultaneously, from equation~(\ref{eq-sepx-equil-endpt}) it then
follows that $B_{\rm pol}^I(l)=B_{\rm pol}^{II}(l)$, and hence no 
finite toroidal current flows along the separatrix either (we make 
a very natural assumption that ${\bf B}_{\rm pol}^I$ is in the same 
direction as ${\bf B}_{\rm pol}^{II}$); thus the separatrix in this 
case is actually not a current sheet; \\
or, \\
{\it (ii)} a finite $I_s\neq 0$ but a divergent poloidal magnetic 
field in the closed-field region, $B_{\rm pol}^I(l) \sim 1/\sqrt
{1-x(l)}\rightarrow\infty$, as $l\rightarrow 0$, $x(l)\rightarrow 1$. 
Note that this situation cannot be dismissed automatically; indeed, 
close to the LC the magnitude of the electric field becomes very 
close to that of the poloidal magnetic field, with the infinitely 
strong electric forces balancing out the infinitely strong magnetic 
ones to a high degree. Then, the separatrix current sheet would have 
to carry back to the star a finite poloidal current~$I_s$ (thus closing 
the poloidal current circuit), and also a toroidal corotation current 
with the surface current density $I_\phi(l)$ divergent near the LC: 
$I_\phi(l)\sim|B_{\rm pol}^I-B_{\rm pol}^{II}|\sim 1/\sqrt{1-x(l)} 
\rightarrow \infty$; there would also have to be a divergent surface 
charge density on the separatrix: $\sigma(l) \sim |E_{\rm pol}^I-
E_{\rm pol}^{II}|\sim 1/\sqrt{1-x(l)}\rightarrow\infty$.

Note that these conclusions, derived from the separatrix equilibrium 
condition near the~LC, are valid regardless of the angle~$\theta_0$ 
between the separatrix and the equator at the intersection point.

In our analysis in the next section we shall assume that there are no 
infinitely strong fields anywhere in the system, and thus shall dismiss 
situation {\it (ii)} ($x_0=1$, $\Omega={\rm const}$, finite $I_s\neq 0$) 
as unphysical. We shall thus concentrate our attention on situation {\it 
(i)} ($I_s=0$). We shall try to analyze the structure of the pulsar 
equation~(\ref{eq-pulsar}) and obtain the asymptotic solution of this 
equation in the vicinity of the intersection point ($x_0=1$, $z_0=0$).


\section{Behavior near the separatrix Y-point: asymptotic analysis}
\label{sec-analysis}

In this section we perform an asymptotic analysis of the pulsar 
equation~(\ref{eq-pulsar}) in the vicinity of the separatrix 
intersection point under the assumption that it is located at 
the~LC, i.e., $x_0=1$, $z_0=0$. Our approach is actually very 
similar to the analysis of the magnetic field near the endpoint 
of a non-relativistic reconnecting current layer, performed 
previously by Uzdensky \& Kulsrud (1997). 

In addition to assuming $x_0=1$, we shall, in this section,
make use of the following two assumptions: \\
1) the intersection point is a finite-angle Y-point, i.e., the separatrix
approaches the equator at an angle~$\theta_0$, $0<\theta_0<\pi/2$, where 
we measure angles from the radial vector lying on the equator and directed 
{\it toward} the star, as shown in Figure~\ref{fig-geometry};\\
2) the poloidal current is fully closed in the open-field region, 
i.e., $I_s=I(\Psi=\Psi_s)=0$; this assumption is motivated by the 
arguments presented in the previous section.

Furthermore, we also assume symmetry with respect to the equator
and therefore consider only the upper half-space, $z\geq 0$.

When considering the $x$-dependent coefficients in equation~(\ref
{eq-pulsar}) in the vicinity of the Y-point, we can, to lowest order 
in $|x-1|$, replace $x$ by~1 everywhere except where it appears in a 
combination like $1-x$; thus we can rewrite this equation as
\begin{equation}
2\xi \, \biggl[ {{\partial^2\Psi}\over{\partial\xi^2}} +
{{\partial^2\Psi}\over{\partial z^2}} \biggr] +
2 \, {{\partial\Psi}\over{\partial\xi}}  = - II'(\Psi) \, ,
\label{eq-pulsar-endpt}
\end{equation}
where we introduced a new coordinate, $\xi\equiv 1-x$.
Instead of using coordinates $(\xi,z)$, it is actually 
more convenient to use polar coordinates $(r,\theta)$ 
defined by
\begin{eqnarray}
\xi &=& 1-x = r \cos\theta \, ,
\label{eq-polar_coords-1}                    \\
z &=& r \sin\theta \, .
\label{eq-polar_coords-2}
\end{eqnarray}

When making the transition to these coordinates, we shall
make use of the following expressions:
\begin{eqnarray}
\biggl({{\partial r}\over{\partial\xi}} \biggr)_z &=&
{\xi\over r} = \cos\theta\, ; \qquad  \qquad 
\biggl({{\partial\theta}\over{\partial\xi}} \biggr)_z =
-{z\over{r^2}} = -{\sin\theta\over r}\, ;  
\label{eq-coord-derivatives-xi}         \\
\biggl({{\partial r}\over{\partial z}} \biggr)_\xi &=&
{z\over r} = \sin\theta\, ; \qquad  \qquad 
\biggl({{\partial\theta}\over{\partial z}} \biggr)_\xi =
{\xi\over{r^2}} = {\cos\theta\over r}\, ;
\label{eq-coord-derivatives-z}
\end{eqnarray}
as well as the usual expression for two-dimensional (2-D) Laplacian
in polar coordinates:
\begin{equation}
{{\partial^2\Psi}\over{\partial\xi^2}} +
{{\partial^2\Psi}\over{\partial z^2}} = 
\Delta_2 \Psi =
{1\over r}\, {\partial\over{\partial r}} 
\biggl( r{\partial\Psi\over{\partial r}} \biggr) +
{1\over r^2}\, {{\partial^2\Psi}\over{\partial\theta^2}} \, .
\label{eq-2D-Laplacian}
\end{equation}

In these coordinates the poloidal magnetic field components
are written as follows (we do not need these expressions now
but will need them later):
\begin{eqnarray}
B_r &=& {1\over{R_{LC}^2}}\, {\partial\Psi\over{r\partial\theta}}\, ,
\label{eq-B_r} \\
B_\theta &=& -\, {1\over{R_{LC}^2}}\, {\partial\Psi\over{\partial r}}\, .
\label{eq-B_theta} 
\end{eqnarray}

Finally, by using expressions 
(\ref{eq-coord-derivatives-xi})--(\ref{eq-2D-Laplacian}), 
we can rewrite the pulsar equation~(\ref{eq-pulsar-endpt}) 
in our polar coordinates (\ref{eq-polar_coords-1})--(\ref
{eq-polar_coords-2}) as
\begin{equation}
\biggl( {{\partial^2\Psi}\over{\partial r^2}}+ {1\over r^2}\, 
{{\partial^2\Psi}\over{\partial\theta^2}}\biggr)\, r\cos\theta+
2\, {{\partial\Psi}\over{\partial r}}\cos\theta-
{{\partial\Psi}\over{r\partial\theta}}\, \sin\theta=
-\, {1\over 2}\, II'(\Psi)\, .
\label{eq-pulsar-polar}
\end{equation}

Now we need to solve this equation separately in region~I 
(region of closed field lines, $0\leq\theta\leq\theta_0$)
and in region~II (region of open field lines, $\theta_0\leq
\theta\leq\pi$) and match the two solutions together at the
separatrix $\theta=\theta_0$. Since in this paper we are 
dealing only with the local structure of the magnetic field
near the Y-point, we can, without any loss of generality,
adopt the convention of counting the poloidal magnetic flux
$\Psi$ from the separatrix $\theta=\theta_0$ (instead of the
usual convention of counting~$\Psi$ from the rotation axis).
Thus, we shall set $\Psi_s=0$ and, correspondingly, $\Psi>0$
in region~I and $\Psi<0$ in region~II.

Very close to the Y-point, i.e. at distances much smaller
than the Light Cylinder radius ($r\ll 1$), the system lacks 
any natural length scale. Hence, we can expect the radial
dependence of~$\Psi$ to be a power law, which enables us 
to make the following {\it self-similar ansatz}:
\begin{eqnarray}
\Psi_I(r,\theta) &\equiv& \Psi(r,\theta\leq \theta_0) =
r^{\alpha_1} f(\theta)\, ,
\label{eq-Psi_I} \\
\Psi_{II}(r,\theta) &\equiv& \Psi(r,\theta\geq \theta_0) =
-r^{\alpha_2} g(\theta)\, ,
\label{eq-Psi_II}
\end{eqnarray}
[we put a minus sign in equation~(\ref{eq-Psi_II}) in order to 
have $g(\theta)\geq 0$.]

Then, the magnetic field components are given by:\\
in region~I:
\begin{eqnarray} 
B_r^I &=& {1\over{R_{LC}^2}}\, r^{\alpha_1-1} f'(\theta) \, , 
\label{eq-Br-I}   \\
B_{\theta}^I &=& -\, {1\over{R_{LC}^2}}\, \alpha_1\, 
r^{\alpha_1-1} f(\theta)\, ;
\label{eq-Btheta-I}
\end{eqnarray}
and in region~II:
\begin{eqnarray} 
B_r^{II} &=& -\, {1\over{R_{LC}^2}}\, r^{\alpha_2-1} g'(\theta) \, , 
\label{eq-Br-II}   \\
B_{\theta}^{II}&=& {1\over{R_{LC}^2}}\, \alpha_2 \, 
r^{\alpha_2-1} g(\theta)\, ;
\label{eq-Btheta-II}
\end{eqnarray}

Notice that the condition that magnetic field does not diverge
near $r=0$ imposes the restriction
\begin{equation}
\alpha_1,\ \alpha_2 \geq 1 \, ,
\label{eq-alpha>1}
\end{equation}
whereas for the case where the poloidal magnetic flux function 
in region~I diverges as $1/\sqrt{r}$ [i.e., case {\it (ii)} in
\S~\ref{sec-Y-point}], we would have $\alpha_1=1/2$.

As for the poloidal current function $I(\Psi)$ that appears on the
RHS of equation~(\ref{eq-pulsar-polar}), it takes very different 
forms in regions~I and~II. In region~I there should be no toroidal 
field, so
\begin{equation}
I(\Psi>0) \equiv 0 \, .
\label{eq-I-regionI}
\end{equation}

In the open-field region~II, $I(\Psi)$ is not zero but does approach 
zero in the limit $\Psi\rightarrow 0$. Since there is no natural 
a priori magnetic field scale in the vicinity of the Y-point, we 
shall again employ a self-similar ansatz, i.e., assume that~$I$ 
scales as a power of~$|\Psi|$ near $\Psi=0$:
\begin{equation}
I(\Psi<0) = q\,  (-\Psi)^\beta \, , \qquad  \beta>0 \, ,
\label{eq-I-regionII}
\end{equation}
so that the right-hand side of equation~(\ref{eq-pulsar-polar}) becomes
\begin{equation}
-\, {1\over 2}\, II'(\Psi<0) = \kappa \, (-\Psi)^{2\beta-1} \, ,
\label{eq-RHS-regionII}
\end{equation}
where
\begin{equation}
\kappa \equiv  {1\over 2}\, \beta q^2 > 0 \, .
\label{eq-kappa-def}
\end{equation}

Upon substituting relationships~(\ref{eq-Psi_I}) and~(\ref{eq-I-regionI}) 
(for the closed-field region) and~(\ref{eq-Psi_II}) and~(\ref
{eq-RHS-regionII}) (for the open-field region) into the main 
equation~(\ref{eq-pulsar-polar}), we obtain two Ordinary Differential 
Equations (ODEs) for the functions~$f(\theta)$ and~$g(\theta)$, along 
with a relationship between the power-law indices~$\alpha_2$ and~$\beta$. 
In particular, in region~I we get a homogeneous linear 2nd-order ODE 
for~$f(\theta)$:
\begin{equation}
f''(\theta)-f'(\theta)\tan\theta+\alpha_1(\alpha_1+1)f(\theta)=0\, ,
\label{eq-pulsar-I}
\end{equation}
which is to be supplemented by two boundary conditions:
\begin{equation}
f'(\theta=0)=0 \, , \qquad \qquad f(\theta_0)=0 \, .
\label{eq-bc-I}
\end{equation}

Similarly, in region~II we get a non-linear 2nd-order ODE for $g(\theta)$:
\begin{equation}
g''(\theta)\cos\theta - g'(\theta)\sin\theta +
\alpha_2(\alpha_2+1)\, g(\theta) \cos\theta = 
-\kappa\, g^{1-{1\over{\alpha_2}}} (\theta) \, ,
\label{eq-pulsar-II}
\end{equation}
together with the relationship 
\begin{equation}
\beta = 1 - {1\over{2\alpha_2}}
\label{eq-beta-alpha2}
\end{equation}
that follows from the requirement that the LHS of equation~(\ref
{eq-pulsar-polar}) scale as the same power of~$r$ as the RHS, i.e., 
$\alpha_2-1=(2\beta-1)\alpha_2$. [At this point, however, we have 
to remark that one cannot a priori exclude the possibility $\beta>
1-1/2\alpha_2$; in such a case the poloidal current would go to zero 
near the separatrix so rapidly that its contribution to the pulsar 
equation would become completely negligible. However, as we shall 
discuss later in this section, it is possible to show that no 
continuous solutions exist in this case.]

The boundary conditions for equation~(\ref{eq-pulsar-II}) are
\begin{equation}
g(\theta_0)=0=g(\theta=\pi) \, .
\label{eq-bc-II}
\end{equation}

Notice that the constant $\kappa$ that appears on the RHS of 
equation~(\ref{eq-pulsar-II}) is in fact unimportant; it just 
sets the scale of variation of the function $g(\theta)$ and, 
hence, the overall scale of the magnetic field strength. Thus, 
we can rescale $\kappa$ away by incorporating it into the solution;
we do this by defining a new variable
\begin{equation}
G(\theta) \equiv \kappa^{-\alpha_2}\, g(\theta) \, .
\label{eq-G-def}
\end{equation}
Then, equation~(\ref{eq-pulsar-II}) becomes
\begin{equation}
G''(\theta)\cos\theta - G'(\theta)\sin\theta +
\alpha_2(\alpha_2+1)\, G(\theta) \cos\theta = 
-\, G^{1-{1\over{\alpha_2}}} (\theta) \, ,
\label{eq-pulsar-II-G}
\end{equation}
with the same boundary conditions
\begin{equation}
G(\theta_0)=0=G(\theta=\pi) \, .
\label{eq-bc-II-G}
\end{equation}

Thus, we see that in our problem the magnetic field structure near 
the Y-point is completely characterized by four finite dimensionless
parameters: the separatrix angle~$\theta_0$ and the three power-law
indices $\alpha_1$, $\alpha_2$, and~$\beta$. Our goal is to obtain
a unique solution of our problem, that is to determine the values 
of these parameters and, simultaneously, determine the functions 
$f(\theta)$ and~$g(\theta)$. In fact, equation~(\ref{eq-beta-alpha2}) 
already gives us one relationship between the power exponents, but we 
still need three more relationships to fix all four parameters. Therefore, 
we shall now discuss the conditions that will help us obtain these three
additional relationships.

The second condition [the first being equation~(\ref{eq-beta-alpha2})] 
that links our dimensionless parameters is rather obvious; it is the 
condition of force-balance across the separatrix $\theta=\theta_0$ 
and it will give us a relationship between~$\alpha_1$ and~$\alpha_2$. 
We have already discussed this condition in \S~\ref{sec-Y-point} [see 
eqs.~(\ref{eq-sepx-equil-1})--(\ref{eq-sepx-equil-endpt})]; with 
$|{\bf E}|=x |{\bf B}_{\rm pol}|$, $I(\Psi>0)=0$, and $I_s=0$, this 
condition can be written simply as 
\begin{equation}
[B_{\rm pol}^2 (r,\theta_0)]^I = [B_{\rm pol}^2 (r,\theta_0)]^{II} \, .
\label{eq-sepx-equil-3}
\end{equation}

Since in our $(r,\theta)$-coordinates the separatrix is a radial
line $\theta=\theta_0$, and since we expect the poloidal magnetic 
field on the two sides of the separatrix to be in the same direction,
condition~(\ref{eq-sepx-equil-3}) simply means that $B_r^I(r,\theta_0)=
B_r^{II}(r,\theta_0)$. Then, using equations (\ref{eq-Br-I})--(\ref
{eq-Btheta-II}), we immediately obtain a second relationship between 
the power exponents:
\begin{equation}
\alpha_2=\alpha_1 \equiv \alpha \, ,
\label{eq-alpha1=alpha2}
\end{equation}
and also a relationship between the normalizations of the functions
$f(\theta)$ and~$g(\theta)$, cast in terms of their derivatives at
$\theta=\theta_0$:
\begin{equation}
f'(\theta_0) = - g'(\theta_0) \, .
\label{eq-sepx-f'=-g'}
\end{equation}

The remaining two relationships between the dimensionless parameters
come from the properties of equations~(\ref{eq-pulsar-I}) and~(\ref
{eq-pulsar-II-G}) themselves and cannot, unfortunately, be written 
out as explicit algebraic equations containing these parameters. 
This fact, however, can prevent us neither from describing and 
discussing the physical and mathematical conditions on which 
these two additional relationships are based, nor from using 
these conditions to obtain the actual unique values of the 
parameters.

First, we shall show that there is a unique one-to-one relationship 
between the separatrix angle $\theta_0$ and the power exponent~$\alpha$.
This relationship comes from analyzing equation~(\ref{eq-pulsar-I}) for 
the closed-field region~I. For any given $\theta_0$ this homogeneous 
linear equation with boundary conditions~(\ref{eq-bc-I}) is an 
eigen-value problem for the coefficient $a\equiv\alpha(\alpha+1)\geq 2$.
It is in fact very easy to see why this must be the case. Indeed, 
the boundary conditions~(\ref{eq-bc-I}) do not give us any scale 
for~$f$; if some function $f(\theta)$ is a solution of the problem
(\ref{eq-pulsar-I})--(\ref{eq-bc-I}), then $Cf(\theta)$ with an 
arbitrary multiplier~$C$ will also be a valid solution. Thus, the 
normalization of~$f(\theta)$ is arbitrary: if, for given~$\theta_0$ 
and~$\alpha$, a non-trivial solution of the problem (\ref{eq-pulsar-I})
--(\ref{eq-bc-I}) exists, then there will be infinitely many such 
solutions; in particular, there will be a solution with $f(0)=1$.
Thus, we can impose an additional boundary condition $f(0)=1$. 
Then, however, we have three boundary conditions for a 2nd-order 
differential equation, which means that for arbitrary~$\theta_0$ 
and~$\alpha$ the system is over-determined. Hence, such a solution 
will exist not for all values of $\theta_0$ and~$\alpha$; the 
condition that it actually does exist gives us a certain relationship 
between~$\theta_0$ and~$\alpha$: for a given $\theta_0$ we would get 
an infinite discrete spectrum of the allowed values of~$\alpha$. 
Of these, however, we are interested only in the lowest one because 
we want a solution without direction reversals of the magnetic field, 
i.e., with $f\geq 0$ everywhere.

\begin{figure}
\plotone{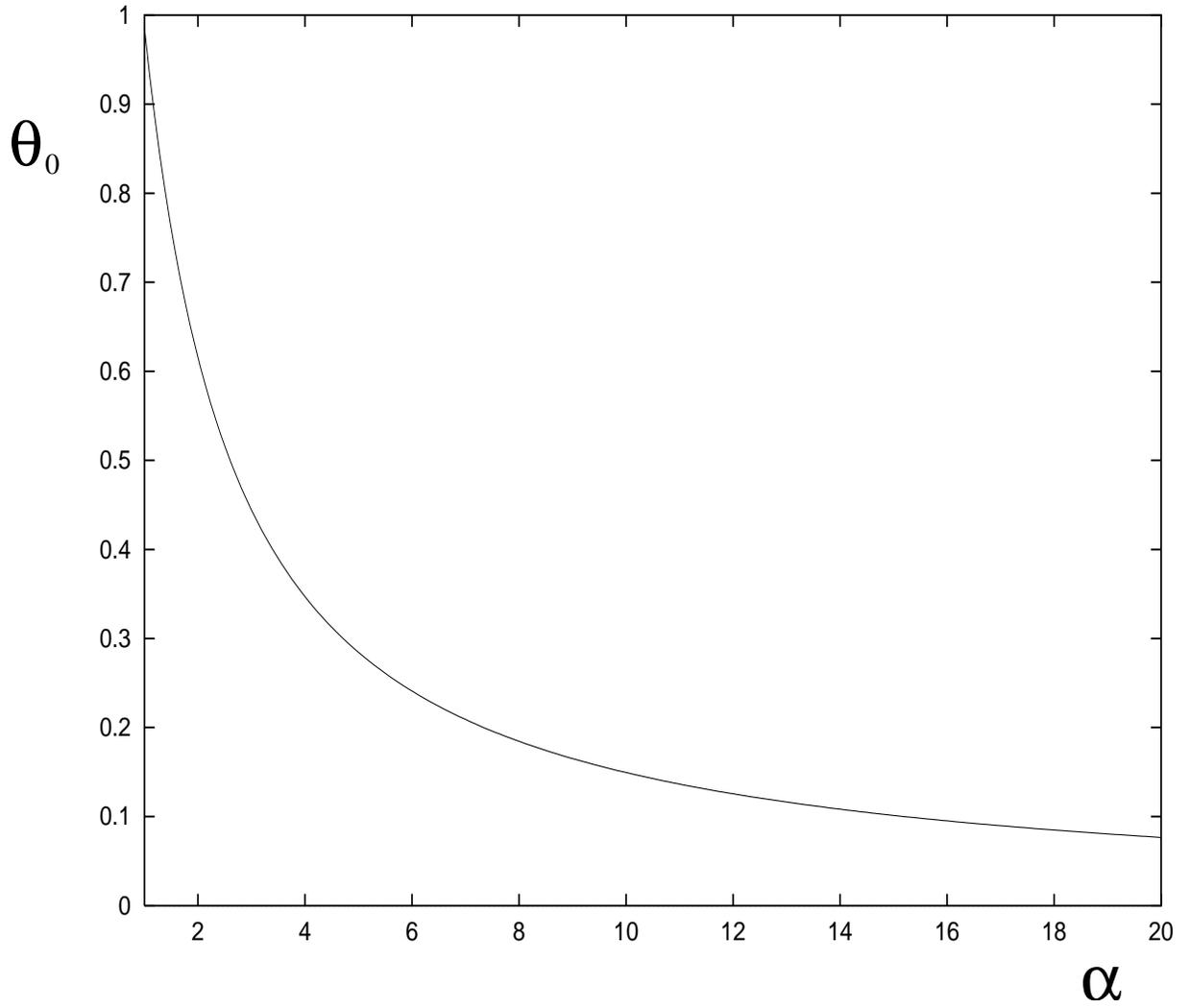}
\figcaption{The separatrix angle~$\theta_0$ as a function of 
the power exponent~$\alpha$.
\label{fig-theta0}}
\end{figure}

\begin{figure}
\plotone{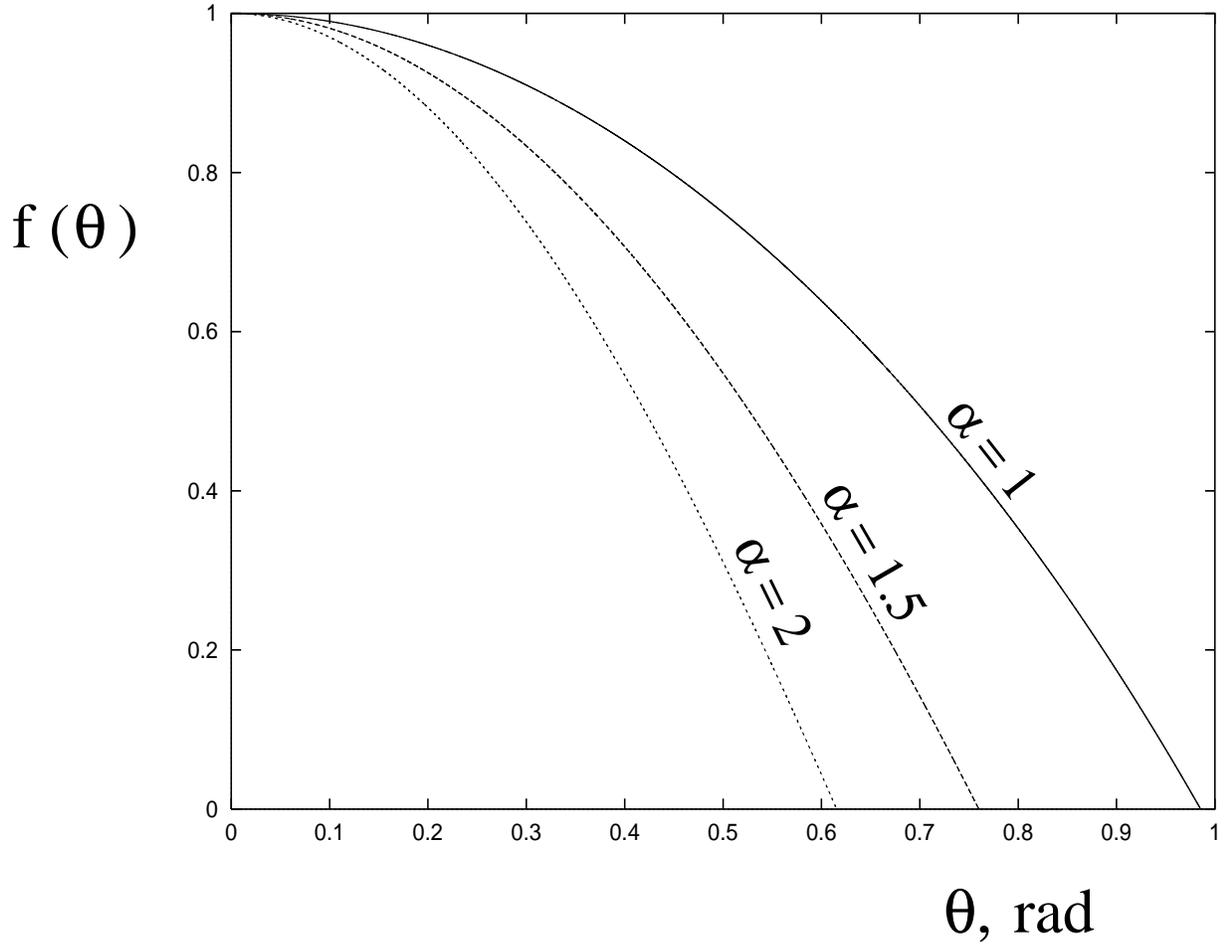}
\figcaption{The function $f(\theta)$ representing the magnetic
flux function in the region of closed-field lines (region~I) for
several selected values of the power exponent~$\alpha$.
\label{fig-f}}
\end{figure}

From the practical point of view, the easiest way to determine
the relationship between~$\alpha$ and~$\theta_0$ is the following.
We scan over the values of~$\alpha$; for each given $\alpha$ we
use our freedom of normalization of~$f(\theta)$ to set $f(0)=1$.
Thus we now have an initial-value problem with two conditions
at $\theta=0$: $f(0)=1$ and $f'(0)=0$. With these two conditions
and with the value of $a\equiv\alpha(\alpha+1)$ given, we can integrate
equation~(\ref{eq-pulsar-I}) forward in~$\theta$ (we do it numerically 
using a 2nd-order Runge--Kutta scheme) until the point where $f(\theta)$ 
becomes zero. This point is then declared the correct value of the 
separatrix angle~$\theta_0$ for the given~$\alpha$. The function 
$\theta_0(\alpha)$ we have obtained as a result of this procedure
is shown in Figure~\ref{fig-theta0}, whereas Figure~\ref{fig-f} 
shows the behavior of the function $f(\theta)$ for several selected 
values of~$\alpha$. We see that $\theta_0(\alpha)$ is a monotonically 
decreasing function with $\theta_0(\alpha=1)=0.986..$ rad [instead of 
$\theta_0(1)=\pi/2$ that one would get if the Y-point were at a finite 
distance inside the LC!], and $\theta_0(\alpha=2)=0.614..\ {\rm rad} 
\approx 35.2^\circ$.%
\footnote
{Note that this value agrees with Michel's (1973a) separatrix 
angle of $\approx 35^\circ$, corresponding to the case of zero 
poloidal current (see also Beskin et al. 1983); this is because 
in his analysis he had in fact assumed that $\alpha=2$ and then 
derived the corresponding value of~$\theta_0$.}
The asymptotic behavior of $\theta_0(\alpha)$ in the limit 
$\alpha\rightarrow\infty$ is $\theta_0\rightarrow\pi/2\alpha$,
which is in fact very easy to obtain analytically.%
\footnote
{Indeed, as $\alpha\rightarrow\infty$, we expect $\theta_0\ll 1$, and 
then we can neglect the $f'\tan\theta$ term in equation~(\ref{eq-pulsar-I}). 
As a result, we get a simple harmonic equation $f''(\theta)+af(\theta)=0$,
with $f'(0)=0$, $f(\theta_0)=0$; the solution of this equation, positive 
everywhere in the domain $0\leq\theta<\theta_0$, is $f(\theta)=
\cos(\pi\theta/2\theta_0)$; it corresponds to $\alpha\simeq\sqrt{a}
=\pi/2\theta_0$.} 
Actually, the exact analytical fit for our numerical curve 
$\theta_0(\alpha)$ appears to be 
\begin{equation}
\theta_0(\alpha) = {\pi\over 2}\, {1\over\sqrt{\alpha(\alpha+1)+b}}\, ,
\label{eq-theta0-alpha}
\end{equation}
with $b\simeq 0.54...$. 

With the dependence $\theta_0(\alpha)$ thus determined, and with
$\beta$ and $\alpha_2$ related to $\alpha$ via equations~(\ref
{eq-beta-alpha2}) and~(\ref{eq-alpha1=alpha2}), respectively, 
we now have everything expressed in terms of~$\alpha$.

The fourth (and final!) condition that will help us determine~$\alpha$
and hence~$\theta_0$ and the rest of the power exponents is the Light
Cylinder regularity condition. This condition states that at the LC
$\theta=\pi/2$ [which is a regular singular point for equation~(\ref
{eq-pulsar-II-G})] the function $G(\theta)$ should be regular, 
namely, it should be a continuously differentiable function, 
with finite 1st and 2nd derivatives. This condition can be 
written as
\begin{equation}
G'\biggl( {\pi\over 2}\biggr) = 
G^{1-{1\over\alpha}}\, \biggl({\pi\over 2}\biggr) \, ,
\label{eq-LC-regularity-G}
\end{equation}
and it is just a particular manifestation of the general LC regularity
condition~(\ref{eq-regularity-LC}). We use this condition to fix the 
unique value of~$\alpha$ (see the general discussion of this issue at 
the end of \S~\ref{sec-basic}). Here is how we do it in practice.

\begin{figure}
\plotone{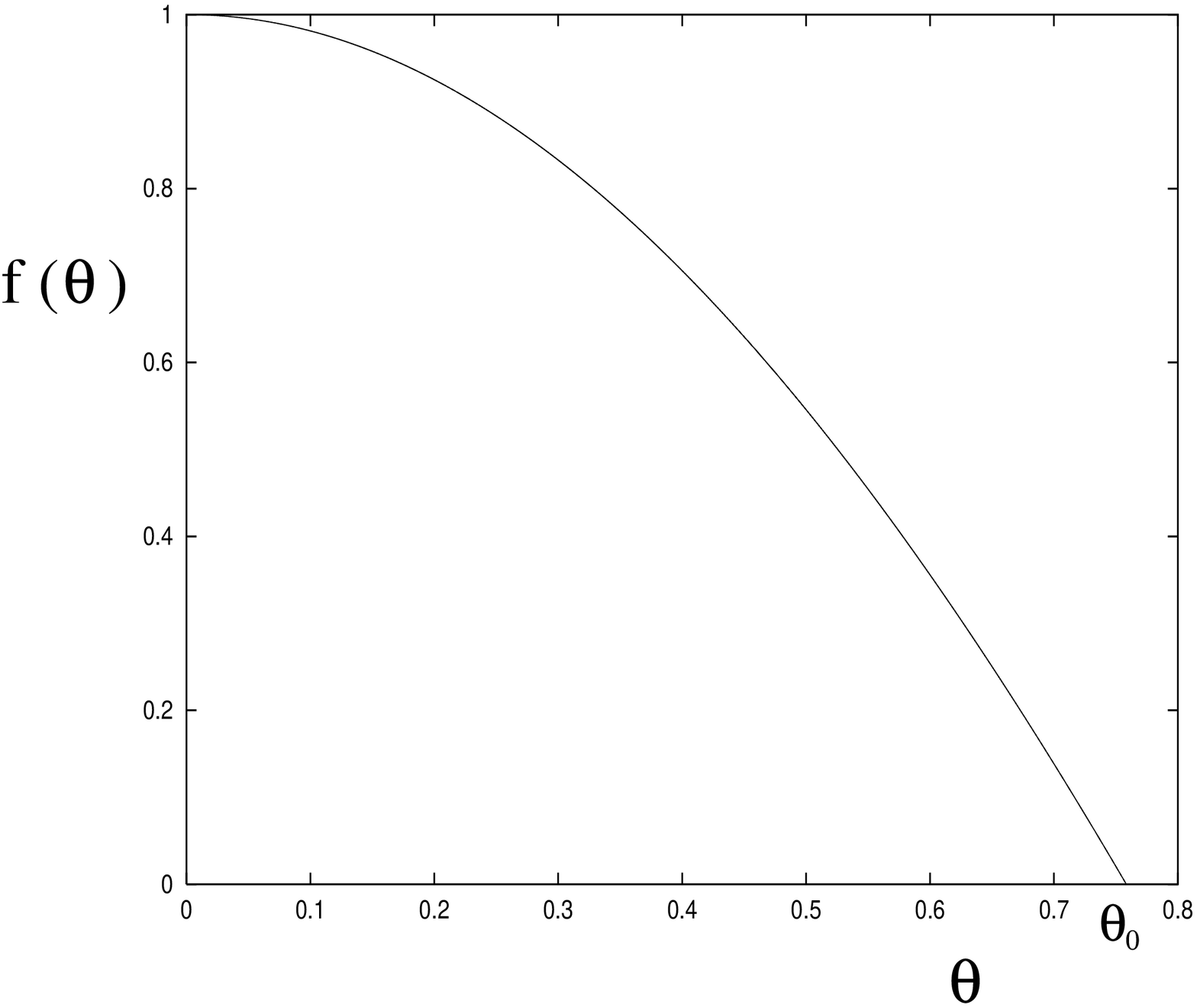}
\figcaption{The function $f(\theta)$ representing the magnetic
flux function in the region of closed field lines (region~I) for
$\alpha=\alpha_0=1.50446...$.
\label{fig-f_0}}
\end{figure}

\begin{figure}
\plotone{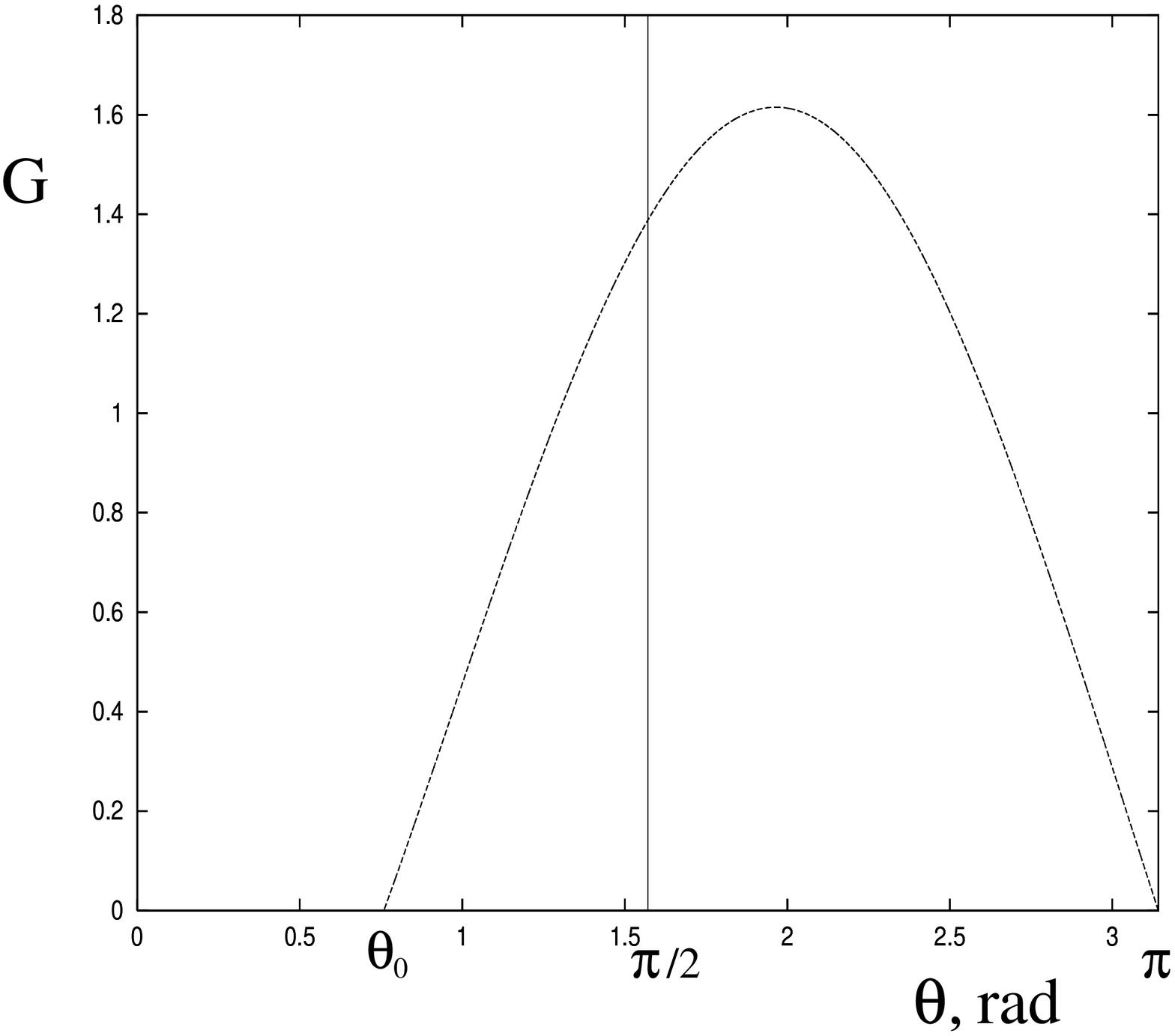}
\figcaption{The function $G(\theta)$ representing the magnetic
flux function in the region of open field lines (region~II) for
$\alpha=\alpha_0=1.50446...$.
\label{fig-G_0}}
\end{figure}

First, we divide up region~II into two sub-regions: 
region~II$^\prime$ (inside the~LC: $\theta_0\leq\theta\leq\pi/2$) and 
region~II$^{\prime\prime}$ (outside the~LC: $\pi/2\leq\theta\leq\pi$). 
Then, we scan over~$\alpha$; for each value of~$\alpha$, we first 
determine the corresponding value of~$\theta_0$ using the procedure 
outlined above. Once $\theta_0$ for a given~$\alpha$ is found, we 
solve equation~(\ref{eq-pulsar-II-G}) separately in regions~II$^\prime$ 
and~II$^{\prime\prime}$ (again, using a numerical shooting method in 
conjunction with the 2nd-order Runge-Kutta integration scheme). In each 
of these regions we use regularity condition~(\ref{eq-LC-regularity-G}) 
at $\theta=\pi/2$ and one of the boundary conditions~(\ref{eq-bc-II-G}) 
at the other end of the region, i.e., $G(\theta_0)=0$ in region~II$^\prime$ 
and $G(\pi)=0$ in region~II$^{\prime\prime}$. Note that the use of the 
regularity condition~(\ref{eq-LC-regularity-G}) guarantees that the 
solutions in {\it each} of the two regions are regular. However, in 
general (i.e., for an arbitrarily chosen~$\alpha$), the two solutions 
obtained in this manner do not match each other at the~LC, the mismatch 
$\Delta G_{\rm LC}\equiv G^{II''}(\pi/2)- G^{II'}(\pi/2)$ [and hence
the mismatch in $G'(\pi/2)$ related to $\Delta G_{\rm LC}$ via eq.~(\ref
{eq-LC-regularity-G})] being $\alpha$-dependent. We find that there is 
only one, special value of~$\alpha$ for which $\Delta G_{\rm LC}$ vanishes 
and the solution continues smoothly across the~LC as one passes from 
region~II$^\prime$ into region~II$^{\prime\prime}$. This special value, 
which we call~$\alpha_0$, is declared the correct value of~$\alpha$. 
Numerically, we find 
\begin{equation}
\alpha=\alpha_0 \approx 1.5045...\, ,
\label{eq-alpha_0}
\end{equation}
and, correspondingly,
\begin{equation}
\beta\approx 0.668...\, , \qquad\qquad \theta_0\approx 0.76 \ {\rm rad}\, .
\label{eq-beta&theta_0}
\end{equation}
This is the way in which the unique correct solution of the problem 
is obtained. The functions $f(\theta)$ and $G(\theta)=\kappa^{-\alpha}
g(\theta)$ corresponding to this value of~$\alpha$ are plotted in 
Figures~\ref{fig-f_0} and~\ref{fig-G_0}, respectively.

Now let us make a little digression and step back to discuss the 
possibility of finding a solution in the case $\beta>1-1/2\alpha_2$.
Recall that in this case the dominant balance of the lowest-order 
(in~$r$) terms in the pulsar equation does not include the contribution 
$-II'(\Psi)/2$ due to the poloidal current. At the same time, note that
relationships~(\ref{eq-alpha1=alpha2}) and~(\ref{eq-sepx-f'=-g'}), 
derived from the separatrix force-balance condition, still hold. 
Hence, the solution in region~II would have to satisfy a linear 
equation that is identical to equation~(\ref{eq-pulsar-I}) for 
$f(\theta)$ on the other side of the separatrix. This fact, combined 
with the condition~(\ref{eq-sepx-f'=-g'}), implies that the solution 
just has to pass smoothly across $\theta=\theta_0$, and thus this point 
[which is just an ordinary point for equation~(\ref{eq-pulsar-I})] does 
not have any special significance; it is just the point where the solution
changes sign. Thus, the general solution in this case takes the form 
$\Psi=r^\alpha f(\theta)$, where $f(\theta)$ satisfies the homogeneous 
linear equation~(\ref{eq-pulsar-I}) in the entire domain $[0,\pi]$ with 
the homogeneous boundary conditions $f'(0)=0$ and $f(\pi)=0$. In addition, 
the solution has to pass smoothly across the regular singular point 
$\theta=\pi/2$ (the Light Cylinder); the regularity condition at
this point is simply $f'(\pi/2)=0$. Such a solution would start
positive with a zero derivative at $\theta=0$, then decrease and
change sign at some angle $\theta_0<\pi/2$, reach a minimum at 
the~LC [since $f'(\pi/2)=0$], and finally increase and go back 
to zero at the equator $theta=\pi$. However, the extra regularity 
condition at $\theta=\pi/2$ makes the system over-determined. Indeed, 
counting two boundary conditions at $\theta=0,\pi$, an arbitrary 
normalization of the solution [e.g., $f(0)=1$], and the LC regularity 
condition, we now have four conditions for our second-order equation 
with one free parameter~$\alpha$. Thus, unless we are somehow extremely 
lucky, the solution of this problem does not exist. One can show that 
this is indeed the case; it turns out that a solution of equation~(\ref
{eq-pulsar-I}) in the region inside the~LC ($[0,\pi/2]$) satisfying both 
the boundary condition $f'(0)=0$ and the LC regularity condition 
$f'(\pi/2)=0$ exists only for even values of $\alpha=2,4,6,...$,
whereas a solution in region ($[\pi/2,\pi]$) satisfying both the 
boundary condition $f(\pi)=0$ and the LC regularity condition exists 
only for odd values of $\alpha=1,3,5,...$. This demonstrates that, in 
the case $\beta>1-1/2\alpha_2$, it is impossible to obtain a solution 
that is continuous at the~LC. This leaves us with the case $\beta=
1-1/2\alpha_2$ (for which we have just found a unique suitable solution)
as the only possibility.

Thus we have managed to obtain a {\it unique} asymptotic solution 
of the ideal-MHD, force-free system (\ref{eq-force-free})--(\ref
{eq-ideal-MHD}) in the vicinity of the separatrix intersection 
point ($x_0=1$, $z_0=0$). This solution, characterized by dimensionless 
parameters~(\ref{eq-alpha_0}) and~(\ref{eq-beta&theta_0}), satisfies the 
separatrix equilibrium condition and is continuous and smooth at the~LC. 
Let us now discuss the physical relevance of our solution to the pulsar 
problem.

Note that the obtained unique solution of our problem has a finite 
value of the derivative $G'(\theta)$ at the equator $\theta=\pi$,
namely, $G'(\pi)=-2.02$, and thus has a non-zero radial magnetic 
field just above (and just below) the equatorial current sheet.
This actually spells bad news for the applicability of our solution.
Indeed, one more condition that needs to be satisfied for the solution
to be physically relevant and that we have not discussed so far is
the condition 
\begin{equation}
|{\bf B}|^2 > |{\bf E}|^2 \, .
\label{eq-B>E-1}
\end{equation}
With $|{\bf E}|$ related to $|{\bf B}_{\rm pol}|$ via equation~(\ref
{eq-|E|}), this condition can be expressed as
\begin{equation}
B_\phi^2 > |{\bf B}_{\rm pol}|^2 \, (x^2-1) \, .
\label{eq-B>E-2}
\end{equation}

We see that this condition is satisfied trivially everywhere inside
the~LC, but is not automatically satisfied outside the~LC. In particular,
in our solution, and in fact in any solution characterized by $I_s=0$,
the toroidal field on the last open field line $\Psi=\Psi_s$ is zero, 
whereas the poloidal magnetic (and hence electric) field is not zero, 
as our explicit solution shows. This means that the inequality~(\ref
{eq-B>E-2}) is violated immediately outside the LC as one moves along 
the equator. More generally, this inequality is violated beyond the 
so-called Light Surface (defined as the surface where $|{\bf B}|=|{\bf E}|$) 
that emanates from the Y-point. It is in fact not difficult to find 
the location of this Light Surface in our solution.

Indeed, according to equations~(\ref{eq-Br-II})--(\ref{eq-Btheta-II}), 
we have
\begin{equation}
B_{\rm pol}^2 = {1\over{R_{LC}^4}}\, r^{2\alpha-2} \, 
[\alpha^2 g^2(\theta) +(g'(\theta))^2]\, ,
\label{eq-Bpol^2}
\end{equation}
and, according to equations~(\ref{eq-B_phi-2}), (\ref{eq-I-regionII}), 
and~(\ref{eq-beta-alpha2}),
\begin{equation}
B_\phi^2 = {1\over{R_{LC}^4}}\, {q^2\over{x^2}} \, (-\Psi)^{2\beta}\simeq
 {1\over{R_{LC}^4}}\, q^2 r^{2\alpha-1}\, g^{2-1/\alpha}(\theta) \, .
\label{eq-Bphi^2}
\end{equation}
Using the result~(\ref{eq-Bpol^2}), we can rewrite this as
\begin{equation}
B_\phi^2 = B_{\rm pol}^2 \, {2r\over\beta} \,  
{{G^{2-1/\alpha}(\theta)}\over{\alpha^2 G^2(\theta)+(G'(\theta))^2}} \, .
\label{eq-Bphi^2-Bpol^2}
\end{equation}

Upon comparing this result with 
\begin{equation}
E^2-B_{\rm pol}^2=(x^2-1)\, B_{\rm pol}^2 = 
-2 r\, \cos\theta\, B_{\rm pol}^2\, ,
\label{eq-E^2-Bpol^2}
\end{equation}
we conclude that the Light Surface $B_\phi^2=E^2-B_{\rm pol}^2$ 
emanates from the Y-point at a finite angle $\theta_{LS}$ which
is determined from the algebraic equation [provided that the solution
$G(\theta)$ is known]:
\begin{equation}
{2\alpha\over{2\alpha-1}}\, G^{2-1/\alpha}(\theta_{LS}) = 
-\, [\alpha^2 G^2(\theta_{LS}) +(G'(\theta_{LS}))^2]\, \cos\theta_{LS} \, .
\label{eq-theta_LS-1}
\end{equation}
For our solution described by the parameters~(\ref{eq-alpha_0})--(\ref
{eq-beta&theta_0}), we find
\begin{equation}
\theta_{LS} \approx 2.07\ {\rm rad}\, ,
\label{eq-theta_LS-2}
\end{equation}
which corresponds to the angle $\pi-\theta_{LS}\approx 62^{\circ}$.

Beyond the LS ($\theta>\theta_{LS}$) the force-free equation~(\ref
{eq-force-free}) is not applicable. The solution we have obtained 
in this section is valid in the region $\theta<\theta_{LS}$, but, 
because the process of obtaining this solution also involved the 
region $\theta>\theta_{LS}$ [e.g., via the boundary condition 
$g(\pi)=0$], our solution may not be the correct solution of 
the overall problem.

In fact, we can make an even more general statement: any relevant to 
our problem finite-field magnetic configuration with the  separatrix 
intersecting the equator at the~LC cannot be almost everywhere (that 
is, again, everywhere except at a finite number of infinitesimally 
thin current sheets) ideal and force-free beyond the~LC! Indeed, as 
we saw previously, if one insists on having the separatrix intersection
point at the~LC, then one has to contend either with having $|B_{\rm pol}|$ 
divergent near $x=1$ or with having $I_s=I(\Psi_s)=0$. The latter case, 
however, is characterized by zero toroidal magnetic field everywhere 
along the last open field line $\Psi=\Psi_s$. Provided that the poloidal 
(radial) magnetic field does not vanish identically near the equatorial 
current sheet, it then follows that $E^2-B_{\rm pol}^2=(x^2-1)B_{\rm pol}^2>
B_\phi^2=0$, and so the force-free equation~(\ref{eq-force-free}) becomes 
inapplicable. And, as our particular solution shows, the condition 
$B^2-E^2>0$ is violated not only on some singular surfaces of measure 
zero, such as the equatorial current sheet (that would be acceptable 
to us at this stage, since we do expect our simple force-free, ideal-MHD 
assumptions to break down there anyway), but in a volume of non-zero 
extent in both poloidal directions, e.g., in the region ($\theta_{LS}
\leq\theta\leq\pi$,$r>0$).


\section{Properties of solutions with the separatrix intersecting 
the equator inside the Light Cylinder}
\label{sec-T-point}

Thus we have to conclude that if one is to keep the hope of finding 
a force-free solution that would be valid all the way up to infinity
(in the limit of vanishing plasma density), or at least to some finite
distance beyond the~LC, then one has to make concessions regarding the 
anticipated location of the separatrix/equator intersection point. In 
other words, the requirement that $B^2>E^2$ outside the LC means that 
$I_s$ must be non-zero and this, in turn, forces one to conclude that 
the intersection point must be located at some finite distance inside 
the~LC, i.e., $x_0<1$. This scenario has actually been the subject of 
several studies in the past 20 years (e.g., Beskin et~al.~1983; 
Lyubarskii~1990; Beskin \& Malyshkin~1998). According to equation~(\ref
{eq-sepx-equil-2}), the corresponding finite-$I_s$ configuration has 
to be characterized by $B_{\rm pol}^I$ staying finite in the limit 
$x\rightarrow x_0$, and, therefore, the separatrix should approach 
the equator at a right angle, $\theta_0=\pi/2$, corresponding to a 
T-point (see Lyubarskii~1990).

Let us now ask what other characteristic features should such a 
solution possess? Here we present a few simple facts that can be 
gleaned immediately by inspecting the force-free equation~(\ref
{eq-pulsar}) and its regularity condition (\ref{eq-regularity-LC}). 
First, since beyond $x=x_0$ the last open field line, $\Psi=\Psi_s$, 
is assumed to lie along the equator~$z=0$, the derivative $\partial\Psi/
\partial x$ should be zero everywhere along this line (for $x>x_0$). 
In particular, we should have $\partial\Psi/\partial x (x=1,z=0)=0$, 
but then it follows from equation~(\ref{eq-regularity-LC}) that 
either $I(\Psi_s)$ or~$I'(\Psi_s)$ has to be zero. Since here we 
are considering the case $I_s\neq 0$, we conclude that $I(\Psi)$ 
has to approach $\Psi_s$ with a zero slope, $I'(\Psi_s)=0$. Thus, 
we see that $I(\Psi)\neq 0$ cannot be a simple linear function, 
unless it is constant everywhere. Next, by substituting the result 
$II'(\Psi_s)=0$, together with $\Psi(x>x_0,z=0)=\Psi_s={\rm const}$, 
into equation~(\ref{eq-pulsar}), we find that $(1-x^2)\partial^2\Psi/
\partial z^2(x,z=0)\equiv 0$ for all $x>x_0$. This means that the 
radial magnetic field also approaches the equator~$z=0$ with a zero 
slope, $\partial B_x/\partial z =0$. 

Now let us see what one can deduce by investigating the regularity
condition~(\ref{eq-regularity-LC}) just above the equatorial
current sheet. According to what we have just established, above
the equator the function $I(\Psi)$ can be expanded as
\begin{equation}
I(\Psi<0) = I_s + q\, (-\Psi)^\beta \, ,
\label{eq-Tpoint-I}
\end{equation}
where $\beta>1$ in order to satisfy the condition $I'(\Psi)\rightarrow 0$,
as $\Psi\rightarrow 0$ (here we again set $\Psi_s=0$ to simplify notation).
Then we find, to lowest order in $|\Psi|$,
\begin{equation}
II'(\Psi<0) = -\, \beta q I_s \, (-\Psi)^{\beta-1} \, .
\label{eq-Tpoint-RHS}
\end{equation}

Now, the radial magnetic field just above the equator is, in general, 
not zero, so we can expand $\Psi(x,z\rightarrow 0)$ to lowest order 
in $z$ as
\begin{equation}
\Psi(x,z) = z\Psi_1(x) + ... \, ,
\label{eq-Tpoint-Psi-z>0}
\end{equation}
where $\Psi_1(x)$ is in fact just the rescaled electric field
above the equator, as can be seen from equations~(\ref{eq-E}) 
and~(\ref{eq-B_R-2}):
\begin{equation}
\Psi_1(x) \equiv {{\partial\Psi}\over{\partial z}} (x,z=0) =
-\, x R_{LC}^2\, B_R(x,z=0) = -\, R_{LC}^2\, E_z(x,z=0) \, .
\label{eq-Tpoint-def-Psi_1}
\end{equation}

Upon substituting equations~(\ref{eq-Tpoint-RHS}) 
and~(\ref{eq-Tpoint-Psi-z>0}) into the regularity 
condition~(\ref{eq-regularity-LC}), we get
\begin{equation}
z \Psi_1'(x=1) = -\, {1\over 2}\, \beta q I_s\, (-\Psi)^{\beta-1} =
 -\, {1\over 2}\, \beta q I_s\, z^{\beta-1}\, [-\Psi_1(1)]^{\beta-1} \, ,
\label{eq-Tpoint-regularity-LC-z>0}
\end{equation}
and, therefore,
\begin{equation}
\beta = 2 \, ,
\label{eq-Tpoint-beta=2}
\end{equation}
unless $\Psi_1'(1)\equiv\partial_x\partial_z\Psi(1,0)=0$ 
[and $\beta>2$ if $\Psi_1'(1)=0$].

We can now use this result to analyze the magnetic field in the 
vicinity of the point $(x_0,0)$. Equation~(\ref{eq-pulsar}) 
in the vicinity of this point can be written as
\begin{equation}
(1-x_0^2)\, \biggl[ {{\partial^2\Psi}\over{\partial\zeta^2}} +
{{\partial^2\Psi}\over{\partial{z^2}}} \biggr] + 
{{1+x_0^2}\over x_0}\, {\partial\Psi\over{\partial\zeta}} =
-\, II'(\Psi) \, ,
\label{eq-Tpoint-pulsar}
\end{equation}
where $\zeta\equiv x_0-x$. Now, just as we did in \S~\ref{sec-analysis},
we can make a transition to the polar coordinates: $\zeta=r\cos\theta$,
$z=r\sin\theta$. In these polar coordinates equation~(\ref{eq-Tpoint-pulsar})
becomes 
\begin{equation}
(1-x_0^2)\, \biggl[ {1\over r}\, {\partial\Psi\over{\partial r}} +
{{\partial^2\Psi}\over{\partial r^2}} +
{1\over r^2}\, {{\partial^2\Psi}\over{\partial\theta^2}} \biggr] + 
{{1+x_0^2}\over x_0}\, \biggl[{\partial\Psi\over{\partial r}}\cos\theta-
{\partial\Psi\over{r\partial\theta}}\sin\theta \biggr] = -\, II'(\Psi) \, ,
\label{eq-Tpoint-pulsar-polar}
\end{equation}

Again, similarly to what we did in \S~\ref{sec-analysis} 
[see eq.~(\ref{eq-Psi_II})], let us assume that the magnetic 
flux function in region~II (the region of open field lines, 
$\theta>\theta_0=\pi/2$) is a power law of~$r$ near the point
($x_0,0$):
\begin{equation}
\Psi(r,\theta) = -\, g(\theta) r^{\alpha_2} \, , \qquad \alpha_2>1\, .
\label{eq-Tpoint-Psi_II}
\end{equation}

Then, using equations~(\ref{eq-Tpoint-RHS}) and~(\ref{eq-Tpoint-beta=2}), 
we can express the RHS of equation~(\ref{eq-Tpoint-pulsar-polar}) as
\begin{equation}
RHS=-\, II'(\Psi) = -\, 2qI_s\Psi =2qI_s\, g(\theta) r^{\alpha_2} \, ,
\label{eq-Tpoint-RHS-polar}
\end{equation}
whereas the LHS of this equation can be expressed as
\begin{eqnarray}
LHS &=& -\, (1-x_0^2)\, r^{\alpha_2-2}\, \bigl[ \alpha_2 g(\theta)+
\alpha_2 (\alpha_2-1) g(\theta) + g''(\theta) \bigr]      \nonumber  \\
&-& \, {{1+x_0^2}\over x_0}\, r^{\alpha_2-1}\, 
\biggl[ \alpha_2 g(\theta)\cos\theta - g'(\theta) \sin\theta \biggr] \, .
\label{eq-Tpoint-LHS-polar}
\end{eqnarray}
In this expression, terms of the lowest order in~$r$ are those
proportional to~$r^{\alpha_2-2}$; at the same time we see from 
equation~(\ref{eq-Tpoint-RHS-polar}) that the RHS of equation~(\ref
{eq-Tpoint-pulsar-polar}) scales with~$r$ as~$r^{\alpha_2}$.
Therefore, we see that in this case of the separatrix intersecting 
the equator inside the~LC, $x_0<1$, the contribution of the RHS is 
completely negligible (as $r\rightarrow 0$).

Thus, the dominant balance in equation~(\ref{eq-Tpoint-pulsar-polar})
dictates that the terms of order~$r^{\alpha_2-2}$ should balance each 
other and hence we get a very simple homogeneous linear ODE for the 
function~$g(\theta)$:
\begin{equation}
g''(\theta) + \alpha_2^2 g(\theta) = 0 \, .
\label{eq-Tpoint-harmonic}
\end{equation}
The boundary conditions for this equation are
\begin{equation}
g(\theta_0=\pi/2) = 0 = g(\pi)\, ,
\label{eq-Tpoint-bc-II}
\end{equation}
and so the solution with the lowest $\alpha_2$ (corresponding to
the simplest magnetic field topology in this region) obviously is
\begin{equation}
g(\theta) = -\, g_{\rm max}\, \sin{2\theta}\, ,
\qquad {\pi\over 2}\leq\theta\leq\pi \, ,
\label{eq-Tpoint-g}
\end{equation}
and, correspondingly, 
\begin{equation}
\alpha_2 = 2 \, .
\label{eq-Tpoint-alpha2=2}
\end{equation}
Thus, $\Psi_{II}(r,\theta)=g_{\rm max}r^2\, \sin{2\theta}$.%
\footnote
{The fact that we have been able to find such a simple non-divergent 
solution near the T-point, whereas Lyubarskii (1990) has found only
a logarithmically-divergent one, can be explained by noting that, 
as we have shown in the beginning of this section, $II'(\Psi_s)$
must be equal to zero according to the LC regularity condition,
whereas Lyubarskii assumed that $I(\Psi)\sim\Psi$ and hence had 
a finite, non-vanishing~$II'(\Psi_s)$.}

As for region~I (the region of closed field lines), the magnetic 
field there is, to lowest order in~$r$, purely vertical and finite 
at $x\rightarrow x_0$, so we can write
\begin{equation}
\Psi_I(r,\theta) = f_{\rm max} \zeta =  f_{\rm max} r\cos\theta \, ,
\label{eq-Tpoint-Psi_I}
\end{equation}
where $f_{\rm max}$ is related to~$I_s$ via the separatrix 
equilibrium condition~(\ref{eq-sepx-equil-2}):
\begin{equation}
f_{\rm max} = {I_s\over{x_0\sqrt{1-x_0^2}}} \, .
\label{eq-Tpoint-sepx-equil}
\end{equation}

We express hope that these simple findings will be helpful in setting 
up or checking the correctness of future numerical attempts to solve 
this nontrivial problem.


\section{Discussion and conclusions}
\label{sec-conclusions}

In this paper we have considered the axisymmetric force-free 
magnetosphere of an aligned rotating magnetic dipole, under 
the additional assumptions of ideal MHD and of the uniformity 
of the field-line angular velocity [$\Omega(\Psi)={\rm const}$]. 
This fundamental model problem is of great importance to any
attempts to understand the workings of radio-pulsars.

More specifically, we have focussed most of our attention on 
the structure of the magnetic field in the vicinity of the point 
of intersection of the separatrix (between the closed- and open-field 
regions) and the equator. We call this point the separatrix intersection
point. The unique singular nature of this point makes it play an extremely 
important role in the overall global problem; in particular, without a 
thorough understanding of the subtleties of the magnetic field behavior 
near this point, it is impossible to prescribe the correct global boundary 
conditions in any sensible way.

We start, however, by discussing (in \S~\ref{sec-basic}) the basic 
equations governing the global force-free pulsar magnetosphere. 
We give special attention to the role played by the Light Cylinder 
regularity condition in determining the poloidal electric current. 
After this general discussion, the rest of the paper is devoted 
entirely to the analysis of the separatrix intersection point.

We first consider the separatrix equilibrium condition in the vicinity 
of this point (see \S~\ref{sec-Y-point}) and find that if it lies at 
the Light Cylinder, then all the poloidal current~$I(\Psi)$ has to 
return back to the pulsar in the open-field region above the equator, 
i.e., there should be no singular current running in a current sheet 
along the equator and the separatrix. We then perform (in \S~\ref
{sec-analysis}) an asymptotic analysis of the relativistic Grad--Shafranov, 
or pulsar, equation in the vicinity of such an intersection point located
at the Light Cylinder. We find a unique self-similar solution that can 
be described by the power law $\Psi\propto r^\alpha$, where $r$ is the 
distance form the intersection point and $\alpha\approx 1.5045$, and 
by the equator--separatrix angle $\theta_0\approx 0.76$ rad. However, 
a further analysis of this solution in the region outside the Light 
Cylinder shows that a Light Surface (a surface where $|B|=|E|$) appears 
just outside the Light Cylinder; in particular, we find that this Light 
Surface originates right at the intersection point and extends outward 
at a finite angle with respect to the equator. The appearance of a Light 
Surface in this case is, of course, not surprising, taking into account 
the fact that (in this case of the separatrix intersection point lying 
at the Light Cylinder) the poloidal electric current~$I$ and hence the 
toroidal magnetic field have to become zero on the last open field line. 
We therefore conclude that the only possibility for an ideal-MHD force-free 
magnetosphere above the putative equatorial current sheet to extend at 
least some finite distance beyond the Light Cylinder, is for the separatrix
intersection point to be located inside, as supposed to right at, the Light 
Cylinder. (Of course we understand that the exact location of this point 
can only be found as a part of a global solution of the pulsar equation 
and that it is impossible to determine it from our local analysis).

These findings give us the motivation to consider the case when
the intersection point lies at some finite distance inside the 
Light Cylinder. In \S~\ref{sec-T-point} we examine the behavior 
of the function $I(\Psi)$ near the last open field line $\Psi=\Psi_s$ 
(above the equatorial current sheet) for this case; we find that the 
derivative $I'(\Psi)$ has to go to zero as $\Psi\rightarrow\Psi_s$, 
whereas the current itself can approach a finite value $I_s\neq 0$. 
We then perform an asymptotic analysis of the magnetic field around 
the intersection point; it is essentially a somewhat simplified and 
trivialized analogue of our analysis in \S~\ref{sec-analysis}. We 
find that the separatrix approaches the equator at a right angle, 
$\theta_0=\pi/2$, and that the field in the region of open field 
lines behaves simply as $\Psi(r,\theta>\pi/2)\sim -r^2\sin{2\theta}$,
while in the closed-field region the magnetic field is simply vertical 
and finite near the intersection point (we call this configuration the 
T-point).

Finally, we would like to make several remarks regarding the 
numerical simulations by CKF and by Ogura \& Kojima (2003):

1) The midplane boundary conditions ($\partial_z\Psi=0$, 
$R<R_{LC}$; $\Psi={\rm const}$, $R>R_{LC}$) adopted by 
both groups have automatically assumed that the separatrix 
intersection point lies at the~LC, and thus have precluded 
them from even considering the possibility $x_0<1$. It is
interesting to note, however, that from the magnetic contour
plots presented by Ogura \& Kojima (2003) it does seem that
the actual intersection point lies a little bit inside the~LC.

2) Neither CKF, nor Ogura \& Kojima (2003) have discussed or even 
mentioned the separatrix equilibrium condition; thus it is not clear 
whether this condition has been satisfied in their simulations. In the 
light of our present findings and of the fact that both of these groups 
have found $I_s\neq 0$, we suspect that the equilibrium condition has 
not, in fact, been satisfied in their studies, at least close to the 
separatrix intersection point. This suspicion is strengthened by the 
fact that both groups have reported having experienced some difficulties 
near the separatrix.

3) Ogura \& Kojima (2003) have reported that they had found $|E|>|B|$ 
at some finite distance outside the~LC, whereas CKF have claimed that 
in their solution $|E|<|B|$ everywhere. The origin of this discrepancy 
is not clear. It may be attributed to the difference in numerical
resolution, although neither of the two groups have conducted very
extensive convergence studies.

I am grateful to Leonid Malyshkin and Anatoly Spitkovsky for 
interesting and insightful discussions. This research was 
supported by the National Science Foundation under Grant No. 
PHY99-07949. 


\section*{REFERENCES}
\parindent 0 pt

Bender, C.~M., \& Orszag, S.~A. 1978, Advanced Mathematical Methods 
for Scientists and Engineers, McGrow-Hill, Inc., New York

Beskin, V.~S. 1997, Phys. Uspekhi, 40, 659 

Beskin, V.~S., Gurevich, A.~V., \& Istomin, Ya.~N. 1983,
Sov. Phys. JETP, 58, 235

Beskin, V.~S., \&  Malyshkin, L.~M. 1998, MNRAS, 298, 847

Contopoulos, I., Kazanas, D., \& Fendt, C. 1999, ApJ, 511, 351 (CKF)

Goldreich P., \& Julian W.~H. 1969, ApJ, 160, 971

Lyubarskii Yu.~E. 1990, Sov. Astron. Lett., 16, 16

Michel F.~C. 1973a, ApJ, 180, 207

Michel F.~C. 1973b, ApJ, 180, L133

Ogura, J., \& Kojima, Y. 2003, astro-ph/0303568

Okamoto, I. 1974, MNRAS, 167, 457

Scharlemann E.~T., \& Wagoner R.~V. 1973, ApJ, 182, 951

Uzdensky, D.~A., \& Kulsrud, R.~M. 1997, Phys. Plasmas, 4, 3960


\clearpage

\clearpage

\clearpage

\clearpage

\clearpage

\end{document}